\documentclass[preprint,12pt]{elsarticle}

\usepackage{graphics}

\usepackage{amssymb}
\usepackage{amsmath}

\journal{Nuclear Instruments and Methods in Physics Research A}

\begin{document}

\begin{frontmatter}



\title{A Modular Path Length Corrector for Recirculating Linacs}


\author[man1]{Hywel Owen\corref{cor1}}\ead{hywel.owen@manchester.ac.uk}
\author[stfc]{Peter Williams}

\cortext[cor1]{Corresponding author. Tel: +44 1925 603797}

\address[man1]{Cockcroft Accelerator Group, School of Physics and Astronomy, University of Manchester, Manchester M13 9PL, UK}

\address[stfc]{STFC Daresbury Laboratory and Cockcroft Institute, Daresbury Science and Innovation Campus, Warrington WA4 4AD, UK}

\begin{abstract}
We present a novel modular magnetic system that can introduce a large and continuously-variable path length difference without simultaneous variation of the longitudinal dispersion. This is achieved by using a combination of an electrically-adjustable magnetic chicane and a mechanically-adjustable focusing chicane. We describe how such a system may be made either isochronous or with a given longitudinal dispersion, and show that the nonlinear terms in such a system are relatively small.
\end{abstract}


\end{frontmatter}



\section{Recirculating Linacs}

Recirculating linac accelerators (RLAs) make use of multiple passes of an electron bunch through the same linac (or set of linacs) to obtain high particle energies \cite{Herminghaus:1991hh}. This principle has been demonstrated in the 6 (soon 12) GeV CEBAF accelerator \cite{Grunder:1988ez,Leemann:2001jh}, and has been proposed both for the acceleration of muons for a neutrino factory \cite{Bogacz:2001gd} and for the generation of high-current, low-emittance beams for electron-proton collisions \cite{Litvinenko:2005uz,Dainton:2008ui}. At sufficiently high beam powers it becomes desirable to recover the energy from the electrons by decelerating them, usually (but not originally \cite{Stein:1985vk,Feldman:1987dh}) in the same structures used for acceleration. The energy recovery linac (ERL) has therefore been proposed for applications where an unrecovered beam power would be prohibitive, either limited by the maximum power that can be directed to a beam dump, or by the more prosaic electrical cost limitation in driving a high power particle beam \cite{Quinn:2005gw,4GLSCDR:2006}.

Presently, the most common proposed application of the ERL is in photon science, either generating low-emittance, high current synchrotron radiation at large electron energies (typically greater than 1~GeV) \cite{Quinn:2005gw,Bondarenko:2007vd,Bilderback:2010ci,Hoffstaetter:2011vr}, or generating high power free-electron laser (FEL) radiation at lower energies (typically 100 to 200~MeV) \cite{Smith:1987wc,Aab:1988hk,AlrutzZiemssen:1991va,Neil:2006cn,Minehara:2006fq}. Beam currents greater than 1~mA are often considered, and sometimes proposed to be as high as 100~mA \cite{Smith:2006uf,Bilderback:2010ci}. Storage rings by comparison are limited by the equilibrium emittance between radiation damping and quantum excitation. Free-electron laser facilities such as FLASH and LCLS that are driven by linacs and which do not use energy recovery are intrinsically limited to lower repetition rates due to their beam power, and which limits the science that may be performed at them; ERL upgrades have been proposed as a way to overcome this \cite{Sekutowicz:2005bx}. Other proposals have considered the use of RLAs to reduce the capital cost required to reach a given electron energy \cite{Williams:2011cb}. Recovery of the energy from the electron bunches is only worthwhile if there is little concomitant power loss in the cavities themselves, and this leads naturally to the use of superconducting cavities in most (but not all \cite{Gavrilov:2007jq}) existing and planned ERLs. 

Both RLAs and ERLs require control of the time of flight (TOF) of successive passes of a particular particle bunch so that the desired phase is seen on each pass with respect to the on-crest value of the cavities. In particular, energy-recovery linacs (ERLs) must carefully control the phase of the returning bunches so that they have approximately a $\pi$ phase difference with respect to the phase seen on the accelerating pass. Accurate phase control is required not just to decelerate the bunches but also to manage instabilities \cite{Herminghaus:1992gj,Merminga:1996bq,Neuffer:1996ux}, the optimal phase difference between accelerating and decelerating passes typically being slightly away from $\pi$ to minimise the final energy spread of the decelerated bunches \cite{Neuffer:1996ux,Neuffer:1997hh}.

Based on the extensive experience at Jefferson Laboratory and from the TESLA linear collider project, the most commonly-considered frequencies for ERL cavities are 1.3 and 1.5~GHz \cite{Gruner:2002gx,Aune:2000ki,Padamsee:2001dw}. The required TOF adjustment required in ERLs is therefore typically no more than 770~ps ($1/f$ at 1.3~GHz, whose wavelength is 23~cm), and this sets the scale for what adjustment systems have to provide.

\section{Magnetic Adjustment - the Bates Bend}

In ERLs that circulate electrons, $\beta\simeq 1$ for all practical purposes, so TOF adjustment by changing particle velocity is not practicable; instead the actual path length that the electrons see is changed. The path length change may be achieved either magnetically or mechanically: both types have been constructed and operated. In the following discussion we use the sign convention that an arc with positive dispersion everywhere has a negative $R_{56}$ term in its total first-order transfer matrix, whereas the common four-dipole chicane used in bunch compressors has a positive $R_{56}$. We label these 'A-type' (arc-like, $R_{56}<0$), and 'B-type' (bunch-compressor-like, $R_{56}>0$).

The prototypical magnetic adjustment design is the so-called 'Bates Bend', which uses a five-dipole arrangement in which a four-dipole chicane is effectively split into two halves by inserting a $\pi$-bend dipole in the middle \cite{Flanz:1981wi,Flanz:1985wm}. The $\pi$ bend has $R_{56}=-L=-R\pi$ (where $\rho$ is the
bending radius). By appropriate choice of bend and separation of the half-chicanes (doglegs) the negative $R_{56}$ from the $\pi$ bend cancels with the positive $R_{56}$ in the doglegs to make the entire system isochronous. Assuming that the dogleg dipoles have the same bend radius as the main dipole, the isochronous condition is obtained when the spacing between the dipoles in each dogleg is

\begin{eqnarray}
\label{bates}
L_D &= &\frac{\rho}{2}  (5-4 \text{cos}[\theta ]+2 \text{cos}[2 \theta ]) \text{cosec}[\theta ] \text{sec}[\theta ]  \\
 & - & \frac{1}{2} \text{cosec}[\theta ]^3 \text{sec}[\theta
] \times \nonumber \\
 &  & \sqrt{\rho ^2 \text{sin}[\theta ]^3 (-2 \text{cos}[\theta ] (\pi +4 \theta -4 \text{sin}[\theta ])+9 \text{sin}[\theta ])} \nonumber
\end{eqnarray}
where $\theta$ is the bend angle in each dogleg dipole, and we assume that the distances between the inner dogleg dipoles and the $\pi$ bend are small. For the special case of $\theta=\pi /4$ this expression reduces to
\begin{equation}
  \label{eq:bates2}
 \frac{L_D}{\rho}=5-2 \sqrt{2}+\frac{\sqrt{8+9 \sqrt{2}-4 \sqrt{2} \pi }}{2^{1/4}}\simeq 0.725.
\end{equation}
In other words, the length/width ratio of a Bates Bend is essentially independent of the bend radius, so to reduce its length the bend radius must also be reduced. In practical designs $\theta$ is set to smaller values nearer to 20$^\circ$ to minimise aberrations \cite{Flanz:1981wi} but a shape scaling still applies.

To adjust the path length in a Bates system a pair of small dipoles each upstream and downstream of the $\pi$ bend provide symmetrical kicks of angle $x^\prime$ to the beam. To understand the effect on the path length we recall the basic equation for path length change through an arbitrary first-order system due to a change in initial position and angle, 
\begin{equation}
  \label{eq:bates3}
\delta l = R_{51} x + R_{52} x^\prime.
\end{equation}
In a dipole magnet of length $L$ and bend angle $\phi$ this becomes
\begin{equation}
  \label{eq:bates4}
\delta l = -\frac{Lx^\prime \cos \phi}{\phi}-x \sin \phi.
\end{equation}
Applying a symmetric condition in the incoming and outgoing $x$ and $x^\prime$ in the dipole gives
\begin{equation}
  \label{eq:bates5}
x= -\frac{Lx^\prime \mathrm{cosec} \phi(1+\cos \phi)}{\phi},
\end{equation}
so that
\begin{equation}
  \label{eq:bates6}
\delta l = -\frac{Lx^\prime}{\phi}.
\end{equation}
These conditions hold for any dipoles, so that in principle bend angles less than $\pi$ may be used to adjust the total path length magnetically, for example by applying an appropriate adjustment of position and angle in the central dipole of a triple-bend achromat (TBA), which in general requires four correctors and which will result in some optical cross-talk. However, in the special case of $\phi=\pi$ in the Bates Bend we have $x=0$: only two correctors are required, and there is little optical cross-talk. Also, the additional aperture required to accommodate the path-adjusted trajectory is only in the central dipole, and not in other elements.

Whilst the Bates Bend scheme is suitable for lower-energy electrons, it is not appropriate for use at the GeV-scale energies needed for shorter-wavelength X-ray production as the $\pi$ bend becomes impracticably large. For example at a field strength of $1.5$~T the $\pi$-bend for 1~GeV electron beam would be 7~m in length.

\begin{figure}[!h]
  \centering
  \includegraphics[width=80mm]{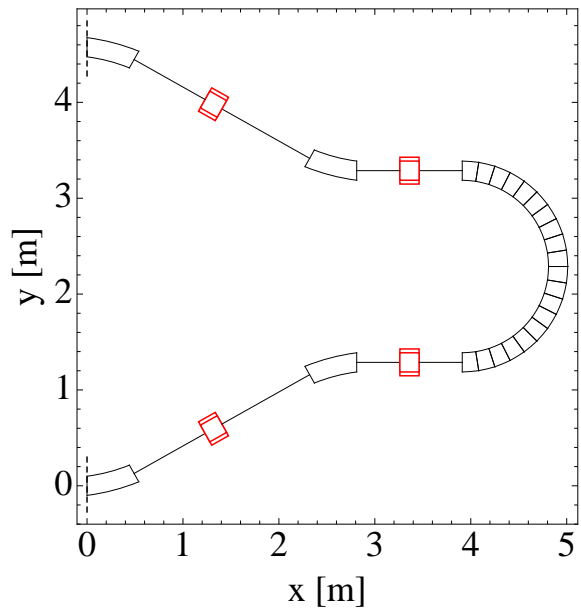}
  \caption{Example Bates arc using $1$~m bend radius dipoles.}
  \label{fig:bates}
\end{figure}

\section{Geometric Adjustment - the Moveable Triple-Bend Achromat}

\begin{figure}[!h]
  \centering
  \includegraphics[width=70mm]{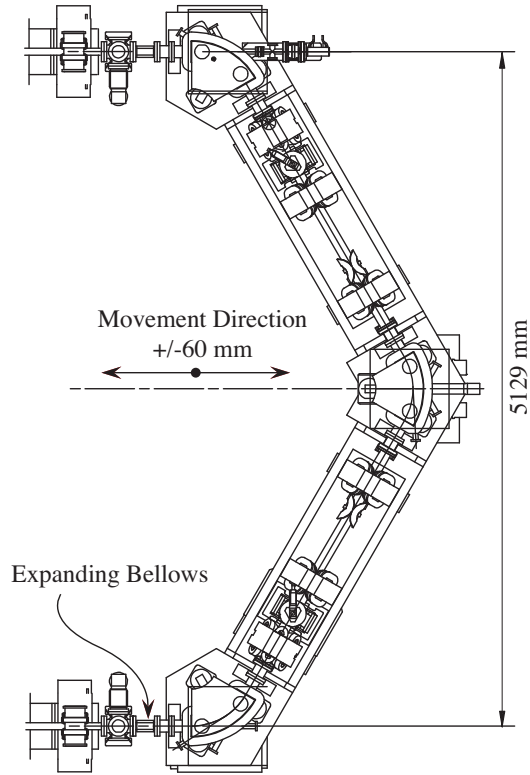}
  \caption{Schematic of the ALICE triple-bend achromat arc, which can isochronously transport an electron bunch of energy up to 35~MeV through 180$^\circ$; the overall arc width is a little over 5~m. Three linear bearings are used with a single motor drive to adjust the longitudinal position of the arc with a measured precision of around 40~$\mu$m, equivalent to a phase accuracy of 0.125$^\circ$.}
  \label{fig:alicearcschematic}
\end{figure}

Guignard~\cite{Letters:vs} showed that a triple-bend achromat (TBA) contains the minimum number
of dipoles necessary for isochronous transport - this is done by driving the
dispersion to a negative value in the central dipole using the intervening quadrupoles. This
method has been used in JAERI~\cite{Minehara:2006fq} and ALICE (formerly ERLP)~\cite{Owen:2004tna}. To produce a variable path length difference, the entire arc must be moved with respect to the rest of the accelerator in a manner analogous to the motion of a trombone. At ALICE we have implemented such a moving arc with good position and therefore phase accuracy: this is shown schematically in Figure \ref{fig:alicearcschematic}. We have measured good longitudinal and angular stability of the arc under motion (with reproducibility of longitudinal position of around 40~$\mu$m at each end of the arc), giving an effective precision of setting the return bunch phase well under the required 1$^\circ$ accuracy needed to obtain efficient energy recovery. In addition, the effect of the expanding expanding drift spaces on the transverse focusing is small, with estimated relative changes of the Twiss functions of order 20\% over the 23~cm range of motion. 

\begin{figure}[!h]
  \centering
  \includegraphics[width=85mm]{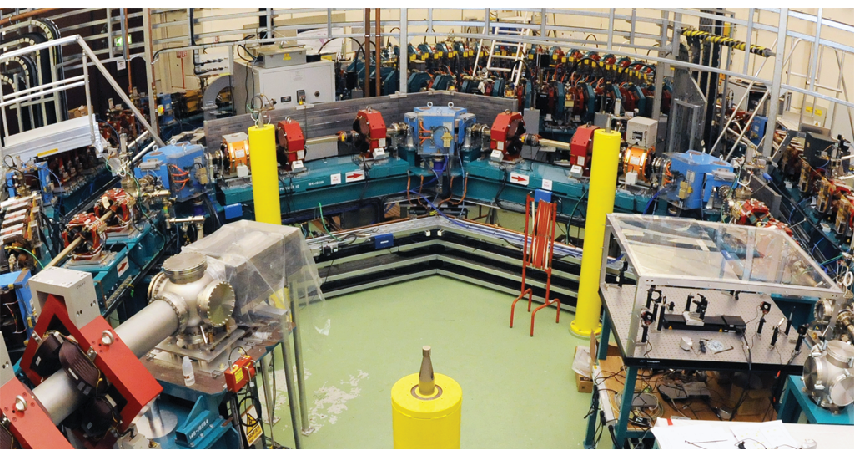}
  \caption{(Colour) Moving TBA arc as installed on ALICE at Daresbury Laboratory. The three blue dipoles each lie over linear bearings upon which the entire arc moves, mounted on a single angled girder.}
  \label{fig:alicearcphoto}
\end{figure}

Whilst good accuracy has been demonstrated for small arcs such as at ALICE, it is impractical to scale this method up to the GeV scale whilst keeping the bending dipoles at reasonable strength, due to the difficulty both of achieving as good position tolerances in, and of moving the increased mass of, a larger arc.

\section{Magnetic Adjustment of High Energy Beams}

At large beam energies the use of both geometric adjustment and the Bates Bend are precluded, and so we need an alternative method. Many ERL projects implicitly assume that path length adjustment is performed using an orbit perturbation in the arcs (see for example discussion in reference \cite{Hoffstaetter:2011vr}); for sufficiently-large arc sizes this perturbation will be small, but there will always be some effect on the transverse optics. We feel this
unnecessarily complicates a recirculating linac design.

Wiggler magnets may be used in situations where longitudinal dispersion change is not as important, for example as proposed in the CLIC CTF3 facility~\cite{Autin:2001wd,Biscari:2000vt} where they propose variable-field, one-period wiggler magnets to introduce a path length change.
This system is compact with a $1.6$~m length, but the path length variation achieved is
small at $\pm 1$~mm for a field variation of $\pm 12$\%.

In a $1.3$~GHz RLA we require a path length change of $\sim 23$~cm. A wiggler $7$~m long with dipoles of variable field from 0~T to 1~T would introduce sufficient change in the path length, but would also generate an $R_{56}$ that varies linearly with the path correction. At
$\delta L=23$~cm, $R_{56}=17$~cm. In principle this can be corrected elsewhere in the lattice \cite{Sharma:2009em}, but this is still a cross-talk that is desirable to avoid.

\section{A Modular Path Length Corrector}

\subsection{\label{sec:fourdipole}The Four-Dipole Chicane}

Given the drawbacks of existing path correction methods presently applied to high-energy electrons, we consider what can be achieved in a modular system. We begin by extending the well-known derivation of the compression provided by a four-dipole chicane, using the hard-edged approximation. We consider four identical dipoles bending in the horizontal plane with no internal focusing and no edge focusing (in other words normal entry and exit from the dipoles).

In terms of the length $L_b$ and bend angle $\theta$ of each dipole, we recall the
dipole transfer matrix is
\begin{equation}
R_b=\left(
\begin{array}{llllll}
 \cos\theta_b  & \frac{L_b \sin\theta }{\theta } & 0 & 0 & 0 & \frac{L_b (1-\cos\theta )}{\theta } \\
 -\frac{\theta_b  \sin\theta }{L_b} & \cos\theta  & 0 & 0 & 0 & \sin\theta  \\
 0 & 0 & 1 & 0 & 0 & 0 \\
 0 & 0 & 0 & 1 & 0 & 0 \\
 -\sin\theta  & -\frac{L_b (1-\cos\theta )}{\theta } & 0 & 0 & 1 & -\frac{L_b (\theta -\sin\theta )}{\theta } \\
 0 & 0 & 0 & 0 & 0 & 1
\end{array}
\right),
\end{equation}
and the drift matrix has its usual form given determined by its length $L_d$. Composing the transfer matrix for the whole chicane is simply
\begin{equation}
R_t=R_b(\theta)\cdot R_d\cdot R_b(-\theta)\cdot R_d\cdot
R_b(-\theta)\cdot R_d\cdot R_b(\theta).
\end{equation}

Explicitly expanding the total longitudinal dispersion $R_t|_{56}$, we have
\begin{eqnarray}
\lefteqn{R_t|_{56}=-4 L_b}\nonumber\\
& & \hspace{0.2in}+\frac{L_b
\sin\theta}{\theta}(-2+11 \cos\theta-6\cos2\theta+\cos3\theta)\nonumber\\
& & \hspace{0.2in}+L_d(-9+16 \cos\theta -5\cos2\theta)
\sin^2\theta\nonumber\\
& & \hspace{0.2in}+\left[\frac{4 L_b}{\theta }-\frac{4 L_d^2 \theta
}{L_b}+\left(-\frac{4 L_b}{\theta }+\frac{6 L_d^2 \theta
}{L_b}\right) \cos\theta \right] \sin^3\theta\nonumber\\
& & \hspace{0.2in}+\left(2 L_d-\frac{L_d^3 \theta ^2}{L_b^2}\right)
\sin^4\theta.
\end{eqnarray}
In our subsequent calculations we use this exact expression, however it is
illustrative to expand it as a power series in $\theta$ to yield
\begin{eqnarray}
\lefteqn{R_t|_{5,6}\simeq\left(\frac{4 L_b}{3}+2 L_d\right) \theta ^2}\nonumber\\
& & \hspace{0.2in}+\left(\frac{23 L_b}{15}+\frac{10 L_d}{3}+\frac{2
L_d^2}{L_b}\right) \theta ^4\nonumber\\
& & \hspace{0.2in}-\left(\frac{1019 L_b}{630}+\frac{206
L_d}{45}+\frac{4 L_d^2}{L_b}+\frac{L_d^3}{L_b^2}\right) \theta
^6\nonumber\\
& & \hspace{0.2in}+O[\theta ]^8
\end{eqnarray}
where the first term is the well-known approximation for small
angles~\cite{Raubenheimer:1993eg}. 

Up to second order in $\theta$, the complete transfer matrix is

\begin{eqnarray}
\lefteqn{R_t\simeq}\nonumber\\
& \left(
\begin{array}{llllll}
R_{11}  & R_{12} & 0 & 0 & 0 & O[\theta ]^3 \\
R_{21} & R_{22} & 0 & 0 & 0 & O[\theta ]^3 \\
 0 & 0 & 1 & 4 L_b+3 L_d & 0 & 0 \\
 0 & 0 & 0 & 1 & 0 & 0 \\
 O[\theta ]^3 & O[\theta ]^3 & 0 & 0 & 1 & R_{56} \\
 0 & 0 & 0 & 0 & 0 & 1
\end{array}
\right) \nonumber\\
&
\end{eqnarray}

where

\begin{eqnarray}
R_{11} & = & 1-\left(8+\frac{6 L_d}{L_b}\right) \theta ^2+O[\theta ]^3  \nonumber\\
R_{12} & = &  \left(4 L_b+3 L_d\right)-\left(\frac{32 L_b}{3}+14 L_d+\frac{4 L_d^2}{L_b}\right)
\theta ^2+O[\theta ]^3  \nonumber\\
R_{21} & = &  -\frac{4 \theta ^2}{L_b}+O[\theta ]^3  \nonumber\\
R_{22} & = &  1-\left(8+\frac{6 L_d}{L_b}\right) \theta ^2+O[\theta ]^3  \nonumber\\
R_{56} & = &  \left(\frac{4 L_b}{3}+2 L_d\right) \theta ^2+O[\theta ]^3  \nonumber\\
\end{eqnarray}

\begin{figure}
\centering
  \includegraphics[scale=0.6]{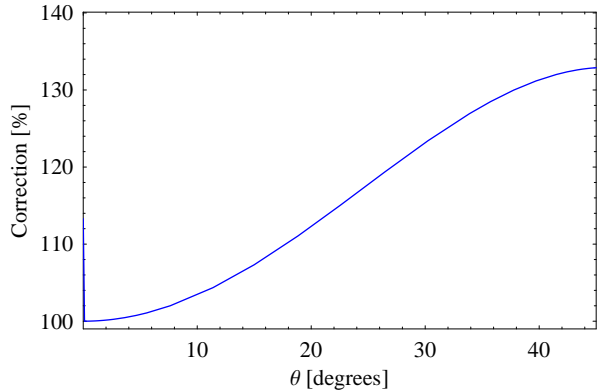}
  \caption{Percentage error associated
    with using the second-order approximation for $R_t|_{56}$ as a function
    of chicane bend angle.}
  \label{fig:seriescorrection}
\end{figure}

The second-order approximation is only accurate for small angles. We illustrate this by considering the limit as $L_d\rightarrow 0$,
\begin{eqnarray}
\lefteqn{\mathop {\lim }\limits_{L_d \to 0 }
\frac{R_t|_{5,6}}{\left(\frac{4
L_b}{3}+2 L_d\right) \theta ^2} =}\nonumber\\
& & \hspace{0.1in}\frac{3}{4 \theta ^3}(4 \sin\theta +4 \sin 2\theta
-4 (\theta +\sin 3\theta)+\sin 4\theta).
\end{eqnarray}
This ratio is shown in
Figure~\ref{fig:seriescorrection}, and shows that the second-order
approximation is accurate to 10\% for bend angles up to $\theta\simeq$ 18$^\circ$; the exact expression for $R_t|_{5,6}$ is therefore needed for bend angles greater than this.

\subsection{\label{sec:dogleg}Four-Dipole Chicane with Focusing}
We next consider a similar four-dipole chicane, except in this case with a pair of quadrupoles inserted into each of the angled sections of the chicane. To suppress dispersion at the centre of the chicane the well-known scheme is used of placing the quadrupoles of each pair $1/4$ and $3/4$ of the way along each angled section, to give a back-to-back pair of doglegs with zero dispersion at each end, and zero dispersion at the centre. The total longitudinal dispersion for this focused chicane may be expressed, using a thin lens approximation for the quadrupoles, as
\begin{equation}
  \label{eq:dogr56}
  \begin{array}{ccl}
    R_t|_{56}=&\frac{1}{f^3 \theta ^2}\left\{ \rule{0pt}{15pt}\right.& 4 L_b^2 
    \left(f^2-L_d^2\right)\sin ^4\theta/2\nonumber\\
    & &-\theta ^2 L_d\sin ^2\theta \left(2 f+L_d\right)
    \left(L_d^2-2 f^2\right)\nonumber\\
    & &\rule{0pt}{15pt}+2 \theta  L_b(\cos\theta-1) L_d\sin\theta \nonumber\\
    & & \hspace{10mm}\times\left(L_d \left(f+L_d\right)-2 f^2\right)\nonumber\\
    & & -2\theta L_b f^3 \left[\theta +(\cos\theta-2) \sin\theta)
    \right]\,\,\left.\rule{0pt}{15pt}\right\},
  \end{array}
\end{equation}
where $f$ is the focal length of each F-quadrupole, $L_b$ is the length of the dipoles and $L_d$ is the drift space length between the dipoles into which each quadrupole pair is placed. With sufficiently-strong quadrupoles the overall longitudinal dispersion may be made negative, in contrast to the conventional chicane where it is always positive. An additional (vertically-focusing) quadrupole is placed at the centre of each dogleg to allow control of the betatron functions through the complete chicane.



\begin{figure*}[htbp]
  \centering
  \includegraphics[width=150mm]{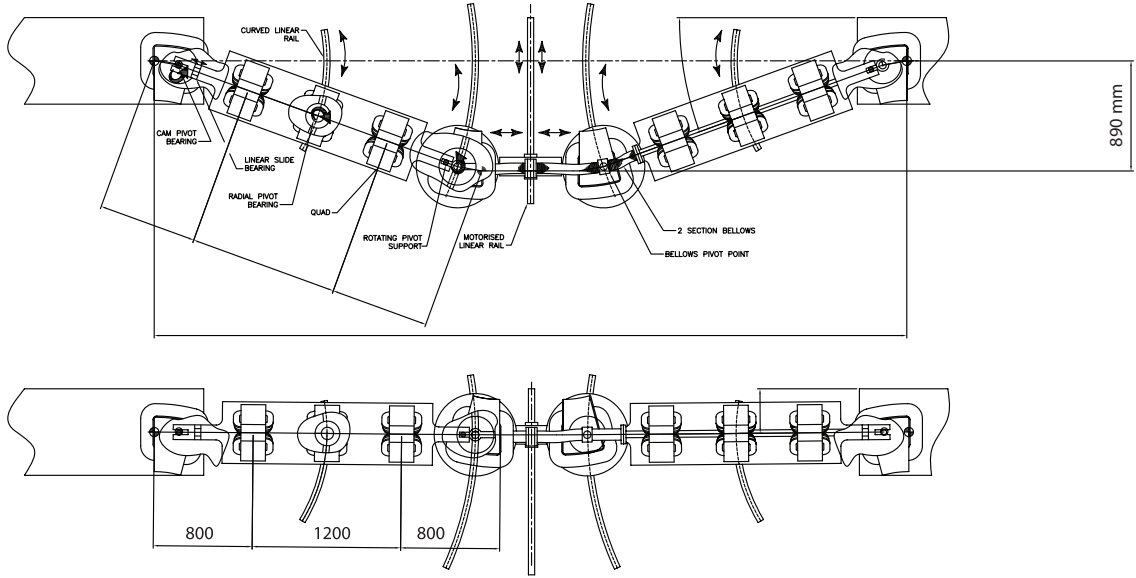}
  \caption{Schematic of an example moveable focusing chicane, showing the dimensions required to introduce a path length change greater than 23~cm ($\lambda$ at 1.3~GHz) using an angle change of 20 degrees (upper figure) from the co-linear closed position (lower figure). The mechanical arrangement is similar in size to the moving ALICE arc described earlier, and smaller than the spin rotator successfully operated at HERA \cite{Buon:1986kx} which also implements a transverse physical shift of a magnetic lattice. Dimensions in the figure are given in mm.}
  \label{fig:pcschematic}
\end{figure*}

\subsection{\label{sec:retard}A Combined Focusing and Non-Focusing Chicane}
  
We now combine the two chicanes - one focusing, one without focusing - into a single path corrector module. The two doglegs from which the focusing chicane is constructed are allowed to move on rails so that the angle of each dogleg may vary from zero degrees upwards, the dipole magnets being adjusted to direct the particle beam at the correct angle. A set of bellows connects the two doglegs together, and expands as the dogleg angles increase, providing directly an increase in the path length the particle beam sees. A schematic of this arrangement is shown in Figure \ref{fig:pcschematic}. Note that the inner focusing chicane dipoles pivot on the girders as the dogleg angle changes, so that the edge angles remain symmetric with the outer dipoles.

Since only a small transverse dispersion $\eta_x$ is generated in the focusing chicane when the dogleg angles increase, the small resultant change in longitudinal dispersion (via the usual $R_{56}=\int \eta_x/\rho_x ds$) it is readily compensated by a smaller electromagnetically-generated change of bend angle in the non-focusing chicane. In the non-focusing chicane a given change in longitudinal dispersion results in a much smaller change in the path length because the horizontal dispersion $\eta_x$ is much larger in the central dipoles than it is in the focusing chicane. The combination of the two chicanes therefore provides a simple way to decouple the longitudinal dispersion from the path length change.

\begin{figure}[!h]
  \centering
  \includegraphics[height=80mm]{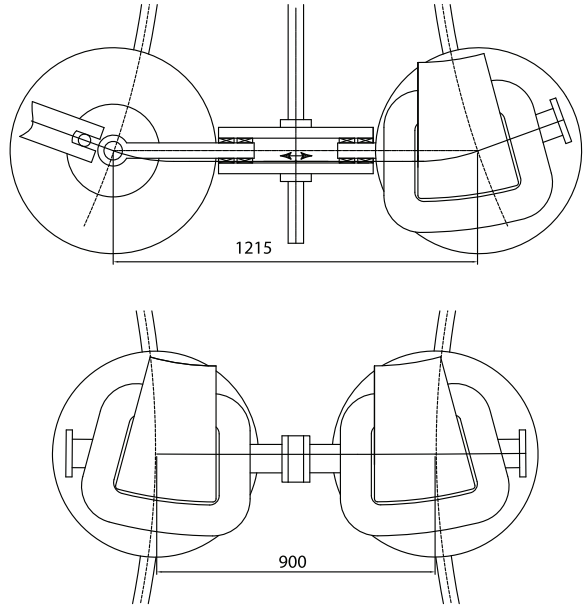}
  \caption{Central region of the focusing chicane, showing the open position (upper figure, showing mechanical guide rods and sleeve) and closed position (lower figure, showing dipoles, vacuum vessel and central bellows); dimensions are shown in mm. An opening angle of 20 degrees provides a change in path length greater than the required 23~cm.}
  \label{fig:pcslide}
\end{figure}

\begin{figure}[tbp]
  \centering
  \includegraphics[height=50mm]{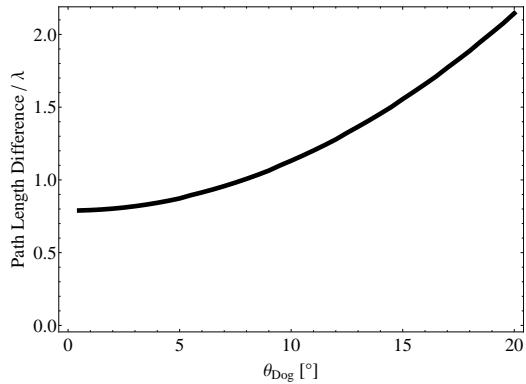}
  \caption{Generated path length (in RF wavelengths) as a function of bend angle in the focusing chicane dogleg arms assuming a non-focusing chicane giving $R_{56}$ of 400~mm, compared to that of a system with no bend at all. 
  the dogleg angle.}
  \label{fig:pld}
\end{figure}

We show as an example a modular path corrector that provides a path length change of $\sim 23$~cm. Figure \ref{fig:pcschematic} shows a mechanical arrangement of a suitable focusing chicane using dipoles of length 20~cm (possible for electron beam energies up to around 200~MeV), whilst Figure \ref{fig:pcslide} shows how the central region expands to provide a physical path length change. The change in path length with bend angle is given in Figure \ref{fig:pld}.

\section{Optics of Example Systems}

\begin{figure}[!h]
  \centering
  \includegraphics[height=45mm]{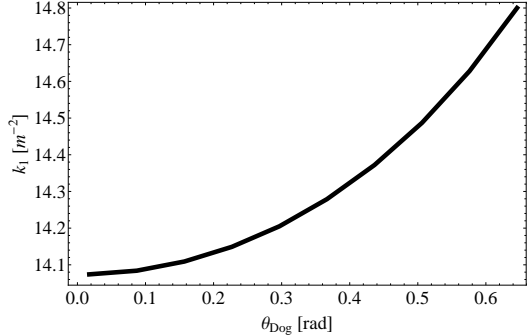}
  \caption{Required strength of the (outer) dispersion-correcting F-quadrupoles as the bend angle changes in the 35/100~MeV focusing chicane layout.
  $L_B=0.2$~m and $L_D=0.5$~m are chosen.}
  \label{fig:kfunc}
\end{figure}

We now consider the beam optics of two example systems that perform path length adjustment, one for low energies (35 or 100~MeV) and one suitable for use on the proposed UK New Light Source (NLS) intermediate recirculation loop at 1200~MeV \cite{Williams:2011cb}. In both cases we assume as before the use of 1.3~GHz superconducting cavities for acceleration. Table \ref{tab:magnettable} give the dipole parameters for the two systems, using the equations given in Sections \ref{sec:fourdipole} and  \ref{sec:dogleg}. The 35/100~MeV layout corresponds to the engineering concept described in Section \ref{sec:retard}, 35~MeV to compare with the ALICE arc design, and 100~MeV as a higher-field design typical of the energies needed for an infrared FEL. Whilst the 1200~MeV layout is longer, the moveable arms of the focusing chicane are still feasible despite having longer dipoles. Figure \ref{fig:kfunc} shows the required focusing strength $k$ of the outer F-quadrupoles in each 35/100~MeV focusing chicane arm which are required to close the dispersion as the bend angle increases; the higher energy layout has similar behaviour.

Our experience on ALICE is that only infrequent movements of the arc are required. As the arc is moved it is advantageous that the transverse focusing is not significantly affected, as this enables operators to tune the path length independently of the optics. Subsequent to the path length change a minor tune-up of the focusing is then performed. Complete decoupling of the optics of the path corrector from the rest of the accelerator is not therefore needed, but the coupling should be small. In the case of the present modular path corrector, we aim that that the path length may be adjusted over a full RF wavelength without overly-large influences in the longitudinal and transverse focusing. 

\begin{table}
\caption{\label{tab:magnettable}.Magnet parameters for the two example path corrector layouts, and used at three illustrative beam energies. FC denotes the focusing chicane dipoles, whilst NC denotes the non-focusing chicane (dogleg) dipoles. At 35 and 100~MeV the layouts are the same, and a representative required $R_{56}$ of 400~mm is assumed, which is a typical requirement at these energies. At 1200~MeV we assume that the path corrector provides the 80~mm $R_{56}$ required in the NLS design at this energy \cite{Williams:2011cb}. Different $R_{56}$ demands may be readily met by adjustment of the NC parameters. The small and changing value of $R_{56}$ generated in the FC section is not shown as it is compensated by a small change in the NC to maintain the overall compression.}
\begin{center}
\begin{tabular}{ccccccc}
\hline
Module & E /MeV & $\theta$ /deg & L /m & B /T & $R_{56}$ /mm \\
\hline
FC & 35 & 20 & 0.2 & 0.207 & - \\
FC & 100 & 20 & 0.2 & 0.585 & - \\
NC & 35 & 21.4 & 0.4 & 0.110 & 400 \\
NC & 100 & 21.4 & 0.4 & 0.313 & 400 \\
\hline
FC & 1200 & 20 & 0.2 & 1.4 & - \\
NC & 1200 & 5.73 & 0.8 & 0.5 & 80 \\
\hline
\end{tabular}
\end{center}
\end{table}

The philosophy of adjusting the path length is to change the FC bend angle (by physically moving the doglegs and compensating the edge angle change with quadrupoles using a look-up table), and simultaneously to make a small magnetic change in the NC to compensate the first-order longitudinal dispersion $R_{56}$. A triplet is used to match between the NC and FC, and of course there will be matching sections into and out of the complete path corrector. Modelling has been performed using either the standard CERN MAD (Methodical Accelerator Design) code \cite{Grote1990}, Elegant \cite{Borland2000}, or a lattice code written in Mathematica that uses equivalent definitions and methods to those of MAD.

\subsection{35/100 MeV Layout}

\begin{figure*}[tbp]
  \centering
  \includegraphics[width=140mm]{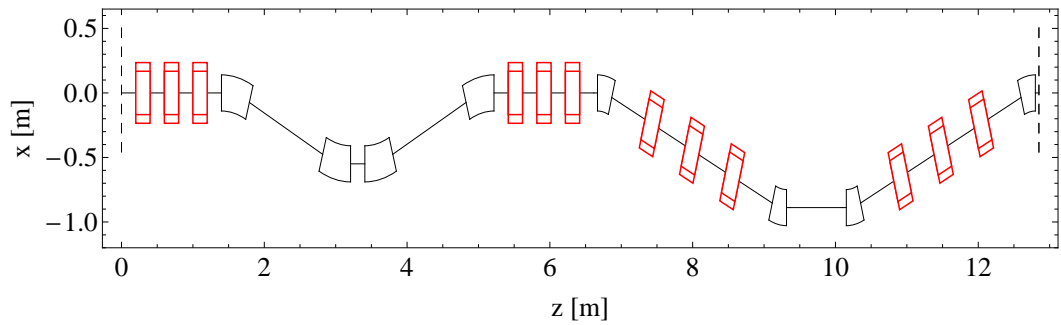}
  \caption{35/100 MeV path corrector in its maximum path length (open) position, with the NC on the left and the FC on the right. Whilst the geometry is correct in this visualisation, the dipole edge angles shown are not to scale.}
  \label{fig:max_draw}
\end{figure*}

\begin{figure}[htbp]
  \centering
  \includegraphics[height=45mm]{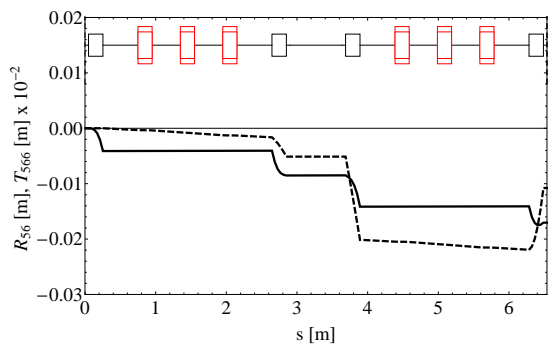}
  \caption{$R_{56}$ (solid) and $T_{566}$ (dashed) in the focusing chicane at maximum path length change (20 degrees FC bend angle).}
  \label{fig:dogleg_r56}
\end{figure}

\begin{figure}[htbp]
  \centering
  \includegraphics[height=45mm]{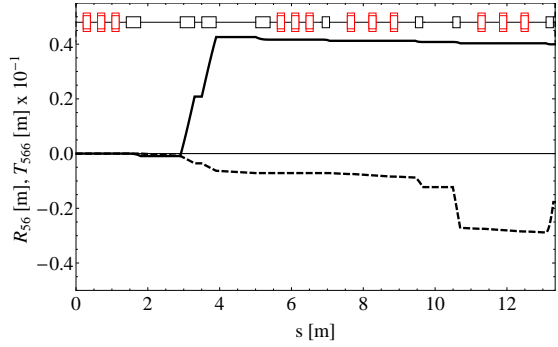}
  \caption{$R_{56}$ (solid) and $T_{566}$ (dashed) in the complete 35/100~MeV path corrector at maximum path length change (20 degrees FC bend angle). The NC bend angle has been changed to compensate the small $R_{56}$ generated by the FC, to maintain an overall value of 400~mm. The $T_{566}$ generated by the FC is similar in magnitude to that generated in the NC, so that linearisation (by arc sextupoles or 3rd-harmonic RF) may still be performed.}
  \label{fig:total_r56}
\end{figure}

\begin{figure}[tbp]
  \centering
  \includegraphics[height=45mm]{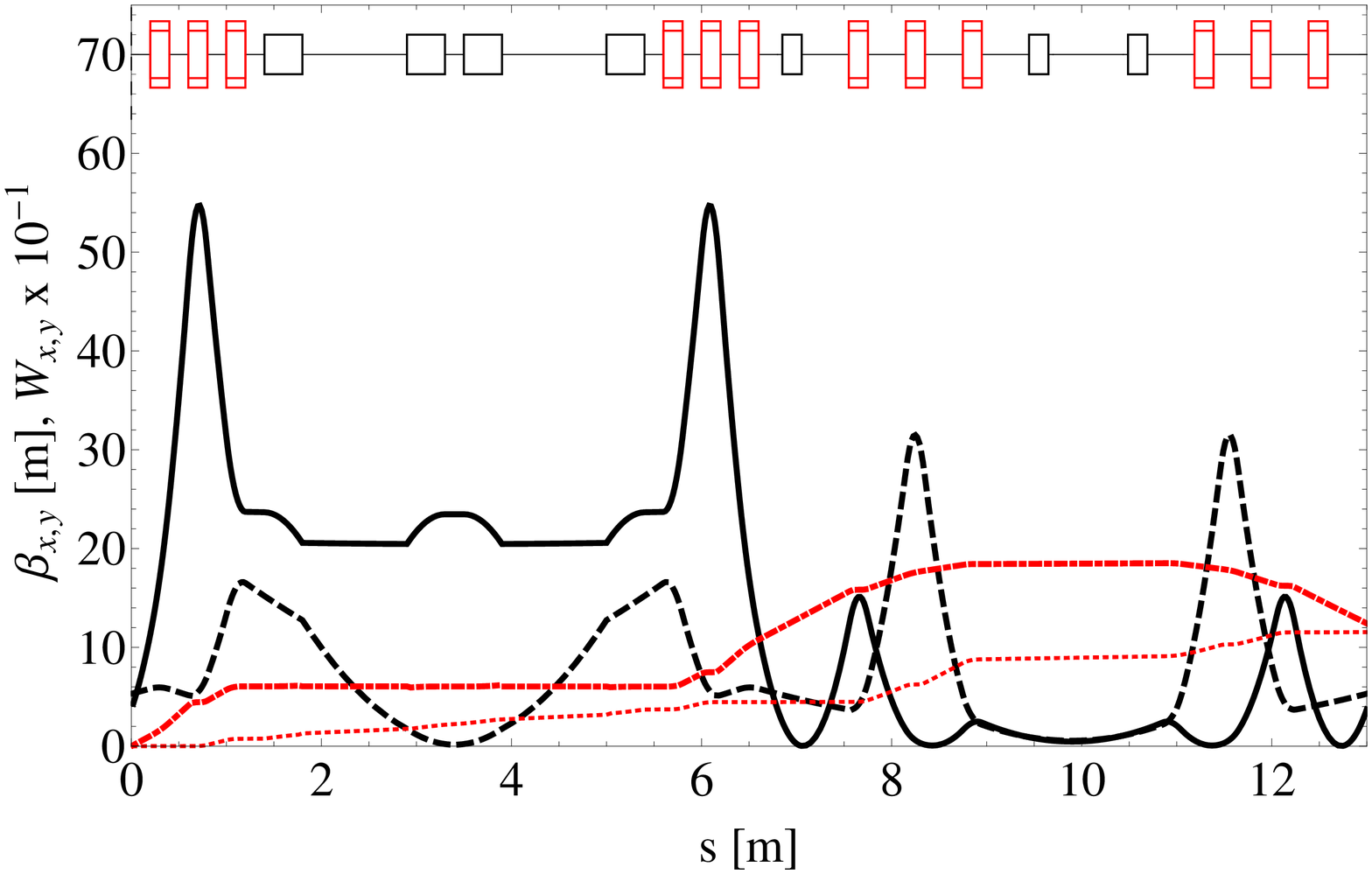}
  \caption{Matched $\beta$ functions in $x$ (solid, black) and $y$ (dashed, black) for the 35/100~MeV path corrector in its minimum path length position, showing also the cumulative chromatic functions $W_x$ (heavy dots, red) and $W_y$ (light dots, red) from $s=0$ at the path corrector entrance.}
  \label{fig:betasmin35}
\end{figure}

\begin{figure}[tbp]
  \centering
  \includegraphics[height=45mm]{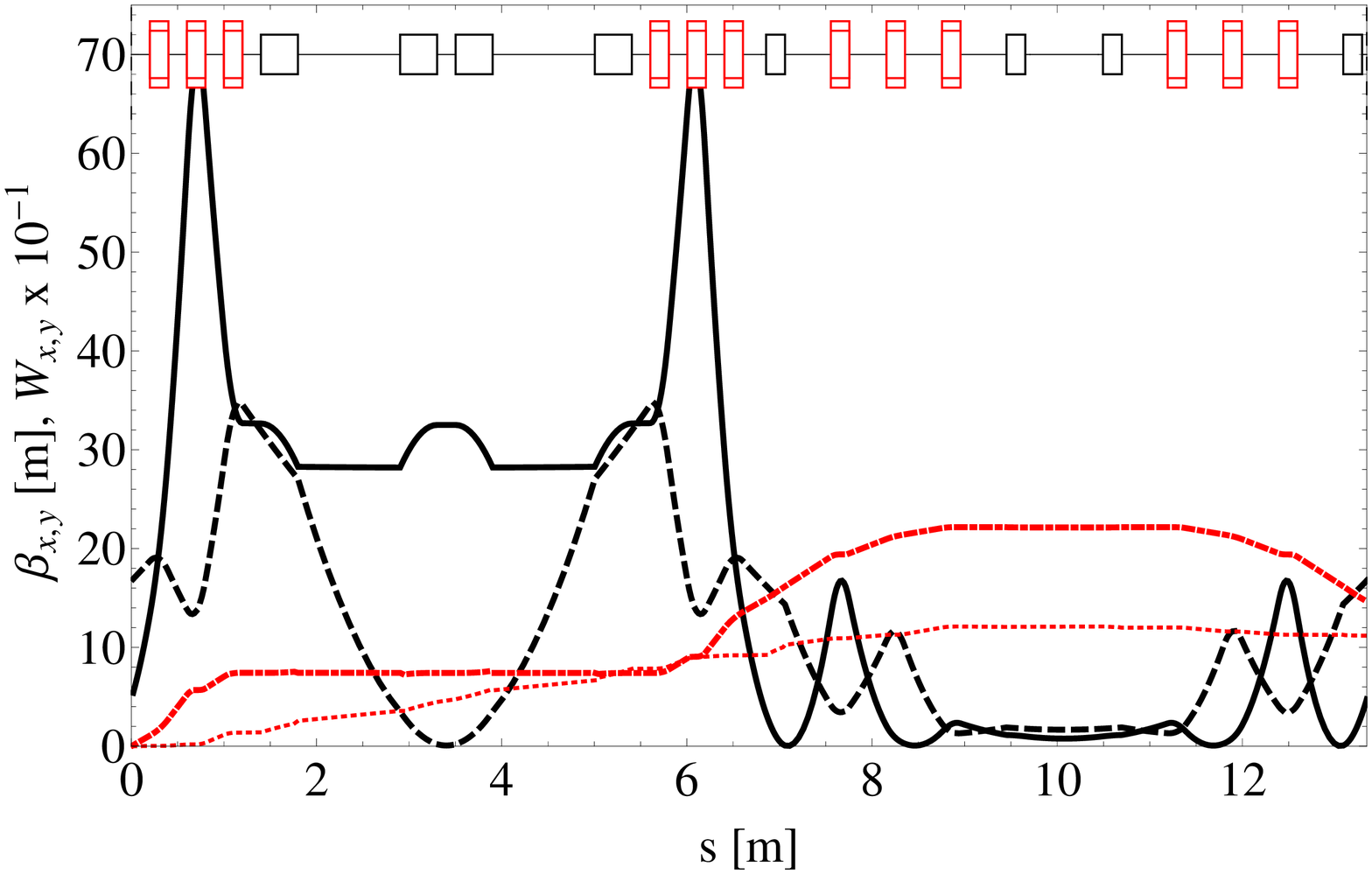}
  \caption{Matched $\beta$ functions in $x$ (solid, black) and $y$ (dashed, black) for the 35/100~MeV path corrector in its maximum path length position, showing also the cumulative chromatic functions $W_x$ (heavy dots, red) and $W_y$ (light dots, red) from $s=0$ at the path corrector entrance.}
  \label{fig:betasmax35}
\end{figure}

\begin{figure}[tbp]
  \centering
  \includegraphics[height=45mm]{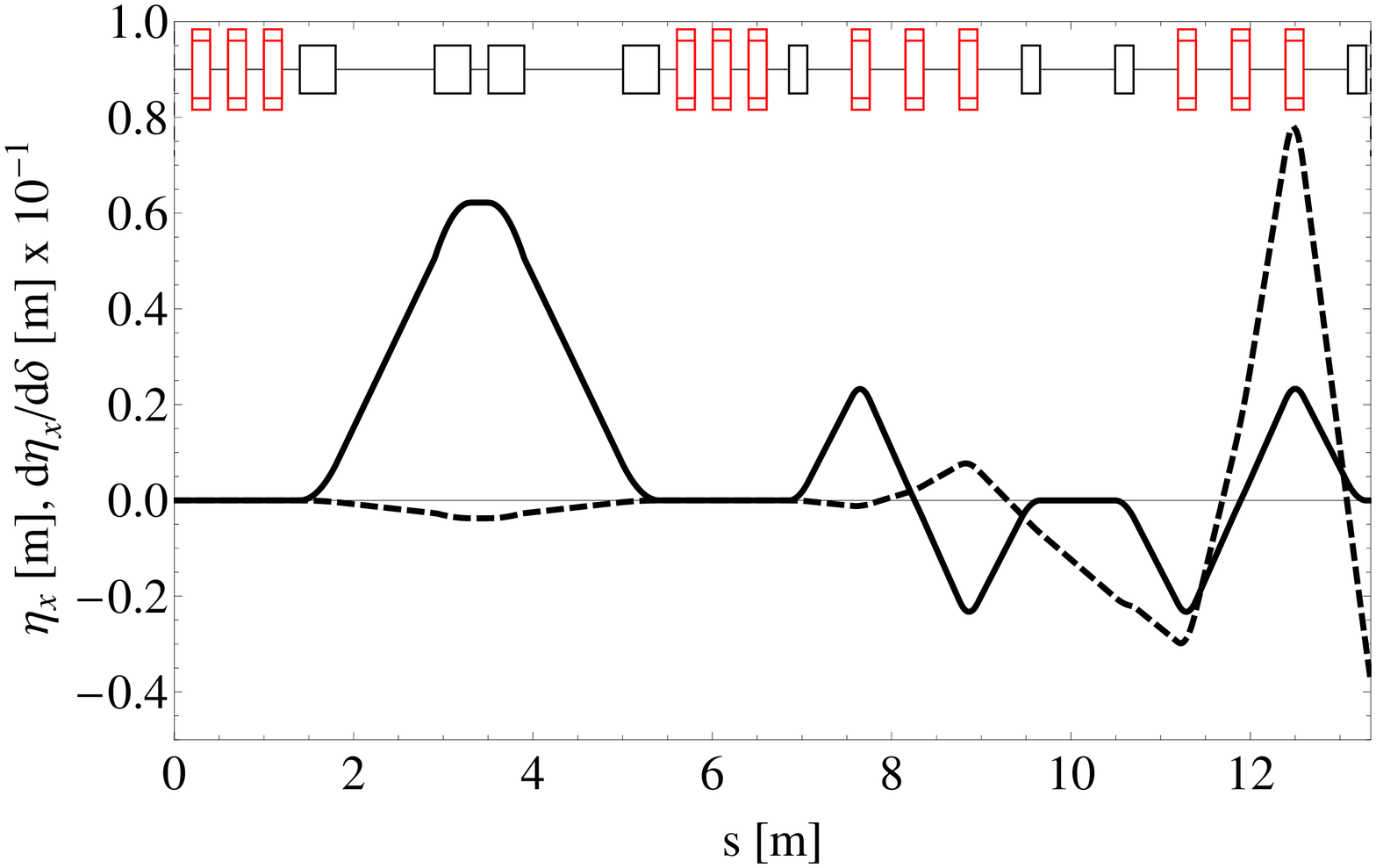}
  \caption{Dispersion $\eta_x$ and chromatic derivative of dispersion $d\eta_x /d\delta$ in the 35/100~MeV path corrector in its maximum path length position. Some beam blowup would be expected from the non-zero $d\eta_x /d\delta$, but this may be zeroed using sextupoles placed adjacent to the FC F-quadrupoles.}
  \label{fig:disp35max}
\end{figure}

Figure \ref{fig:max_draw} shows the 35/100~MeV path corrector layout in its maximum correction positions, corresponding to the engineering layout shown in Figure \ref{fig:pcschematic}.The development of $R_{56}$ and $T_{566}$ is shown in Figure \ref{fig:dogleg_r56} for the focusing chicane alone, and in Figure \ref{fig:total_r56} for the complete path corrector. It may be seen that the additional longitudinal terms introduced by the focusing chicane are similar enough in size that they can be managed in a complete bunch compression system. The matched Twiss functions in the complete path corrector are shown in Figures \ref{fig:betasmin35} and \ref{fig:betasmax35} together with the chromatic amplitudes $W_{x,y}$ defined as
\begin{equation}
  \label{eq:chromatic}
W_{i}=\sqrt{\left (\frac{d\alpha_i}{d\delta}-\frac{\alpha_i}{\beta_i}\frac{d\beta_i}{d\delta}\right )^2+\left ( \frac{1}{\beta_i}\frac{d\beta_i}{d\delta} \right )}\textrm{   }, i=x,y.
\end{equation}
The chromatic amplitudes are reasonable in size: importantly, the contribution to the chromatic amplitudes in the FC is similar in size to that in the NC. The Twiss functions do not change significantly as the FC doglegs are opened, confirming that the longitudinal/transverse coupling is weak and that the path length may be changed without significant disruption to the transverse focusing. The dispersion and chromatic derivative of dispersion $d\eta_x/d\delta$ are shown in Figure \ref{fig:disp35max}, the latter being small enough that it may be readily corrected using small sextupoles placed in the FC.

\subsection{1200 MeV Layout}

\begin{figure*}[tbp]
  \centering
  \includegraphics[width=165mm]{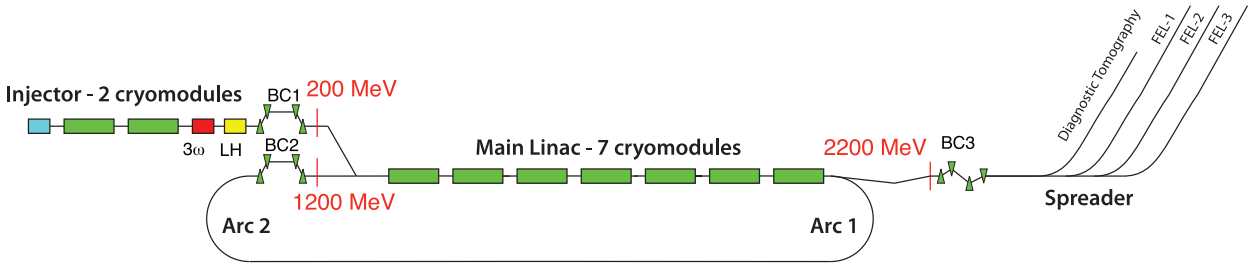}
  \caption{Schematic layout of the recirculating option of the New Light Source, showing the gun (blue), 1.3~GHz accelerating cryomodules (green), third-harmonic (3$\omega$) RF lineariser (red), laser heater (yellow), and the three bunch compressors BC1, BC2 and BC3. The 1200~MeV path corrector performs the tasks both of providing the c.~80~mm $R_{56}$ at BC2, and of providing the necessary path length adjustment.}
  \label{fig:nlslayout}
\end{figure*}

The 1200~MeV layout assumes that it will also serve to provide the $R_{56}$ of 80~mm required in the second compression stage of the recirculating linac option for the proposed UK New Light Source (described fully in Williams et al. \cite{Williams:2011cb}). The recirculating option proposes two passes through the same 1~GeV accelerating section, requiring there to be path length adjustment at 1200~MeV. The proposed layout is shown in Figure \ref{fig:nlslayout}, and the optimised compression parameters are given in Table \ref{tab:nlscompression}. Despite the addition of the four path corrector dipoles in the BC2 section (Bunch Compressor 2), the estimated emittance growth from incoherent and coherent radiation is small. Although a C-type compressor is shown in this study, BC2 may be either C-type or S-type: compensation of the FC $R_{56}$ change may be accomplished in either type. Optimisation of the compression results in actual energies after the first and second passes that are several percent different from the nominal 1200 and 2200~MeV values; similarly, the exact $R_{56}$ values used in the bunch compressors will vary slightly with machine configuration. We have performed calculations here for the nominal $R_{56}=$80~mm.

\begin{table}
\caption{\label{tab:nlscompression}Compression parameters in the recirculating option of the New Light Source, showing how the bunch compressors and RF phases are used to vary the overall compression whilst leaving the fixed sections unchanged; other optical modules in the accelerator have small $R_{56}$ values. Note that the sign convention adopted here is the opposite to that used in Williams et al. \cite{Williams:2011cb}.}
\begin{center}
\begin{tabular}{lcr}
\hline
\textbf{Optimised Sections} & \textbf{Variable} & \textbf{Value} \\
\hline
$3\omega$ Lineariser & $E$ & $16$~MV/m\\
      $3\omega$ Lineariser & $\phi$ & $+160.9$$^{\circ}$\\
      Injector & $\phi$ & $-29.1$$^{\circ}$\\
      Linac First Pass & $\phi$ & $-11.8$$^{\circ}$\\
      Linac Second Pass & $\phi$ & $+2.3$$^{\circ}$\\
      Bunch Compressor 1 & $R_{56}$ & $+96.7$~mm\\
      Bunch Compressor 2 & $R_{56}$ & $+80.9$~mm\\
      Bunch Compressor 3 & $R_{56}$ & $+23.9$~mm\\
\hline
\textbf{Fixed Sections} & & \\
\hline
      Laser Heater Chicane & $R_{56}$ & $+4.9$~mm\\
      Each Arc & $R_{56}$ & $+18.65$~mm\\
\hline
\end{tabular}
\end{center}
\end{table}

\begin{table}[htb]
  \caption{\label{tab:nlsbunch}Bunch length (full width) and relative energy spread (full width) at successive points in the optimised NLS beam transport.}
  \begin{center}
    \begin{tabular}{lrrr}
       \hline
       \textbf{Location} & \textbf{Energy /MeV} & $\sigma_t$ /ps &  $\sigma_E$\\
      \hline
      End of Cryomodule 1 & 136 & 18 & 2.8~\% \\
      After BC1 & 228 & 5.0 & 3.5~\% \\
      Matching to Arc 1 & 1227 & 6.0 & 1.5~\% \\
      After BC2 & 1227 & 0.90 & 1.5~\% \\
      After BC3 & 2244 & 0.35 & 0.6~\% \\
       \hline
    \end{tabular}
\end{center}
\end{table}

\begin{figure}[tbp]
  \centering
  \includegraphics[height=45mm]{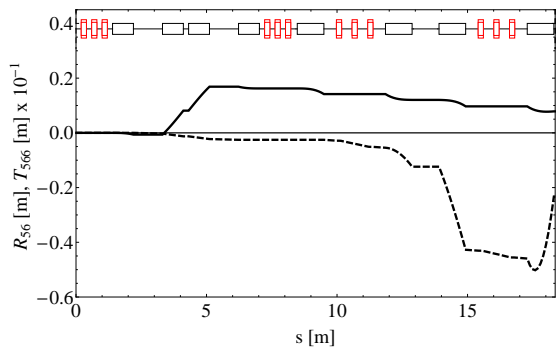}
  \caption{$R_{56}$ (solid) and $T_{566}$ (dashed) in the complete 1200~MeV path corrector at maximum path length change (20 degrees FC bend angle). The NC bend angle has been changed to compensate the small $R_{56}$ generated by the FC, to maintain an overall value of 80~mm. The $T_{566}$ generated by the FC is larger in comparison to the NC contribution than in the 35/100~MeV case, but linearisation may still be carried out using the third-harmonic RF.}
  \label{fig:r561200max}
\end{figure}

\begin{figure}[tbp]
  \centering
  \includegraphics[height=45mm]{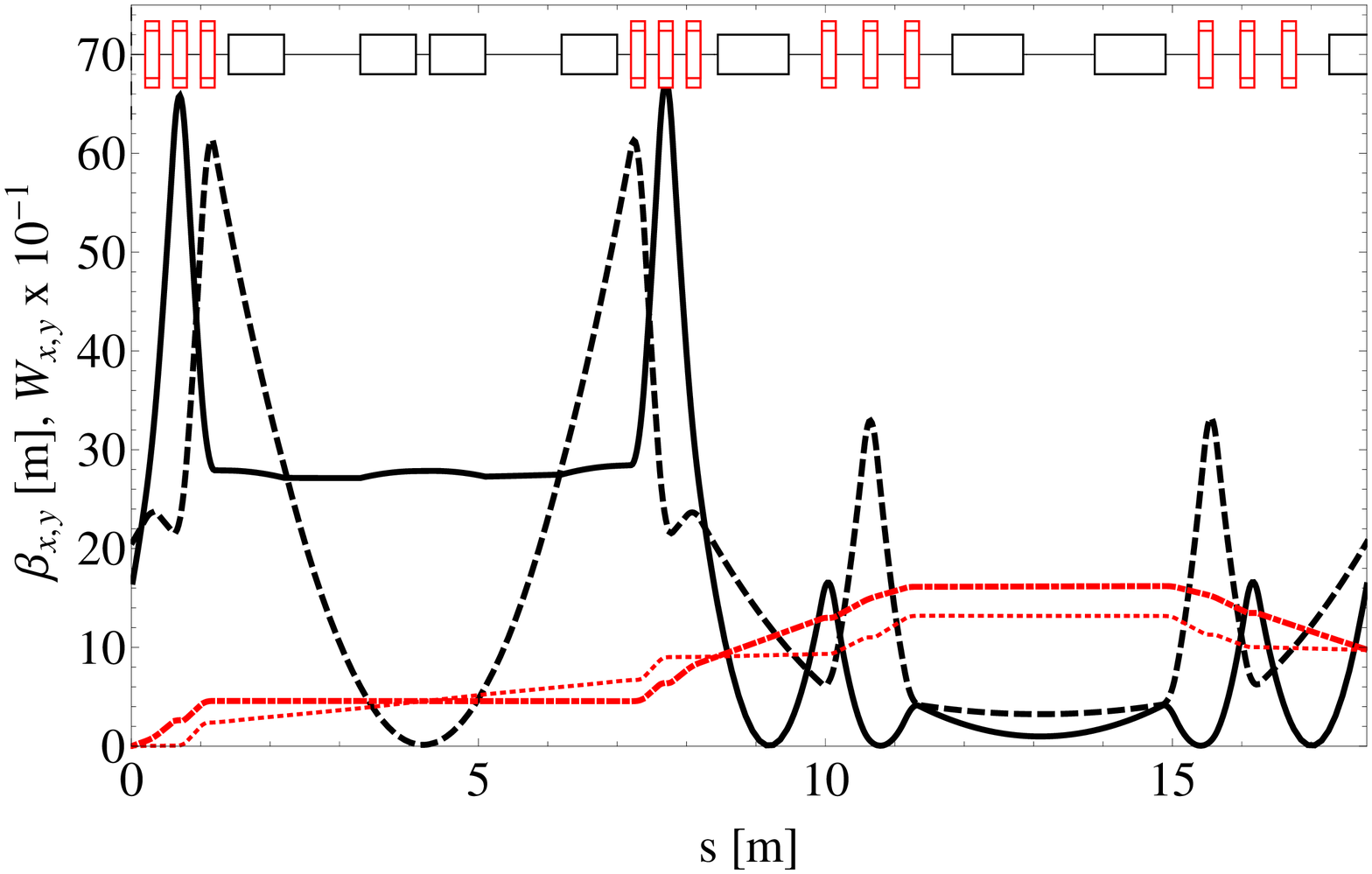}
  \caption{Matched $\beta$ functions in $x$ (solid, black) and $y$ (dashed, black) for the 1200~MeV path corrector in its minimum path length position, showing also the cumulative chromatic functions $W_x$ (heavy dots, red) and $W_y$ (light dots, red) from $s=0$ at the path corrector entrance.}
  \label{fig:betasmin1200}
\end{figure}

\begin{figure}[tbp]
  \centering
  \includegraphics[height=45mm]{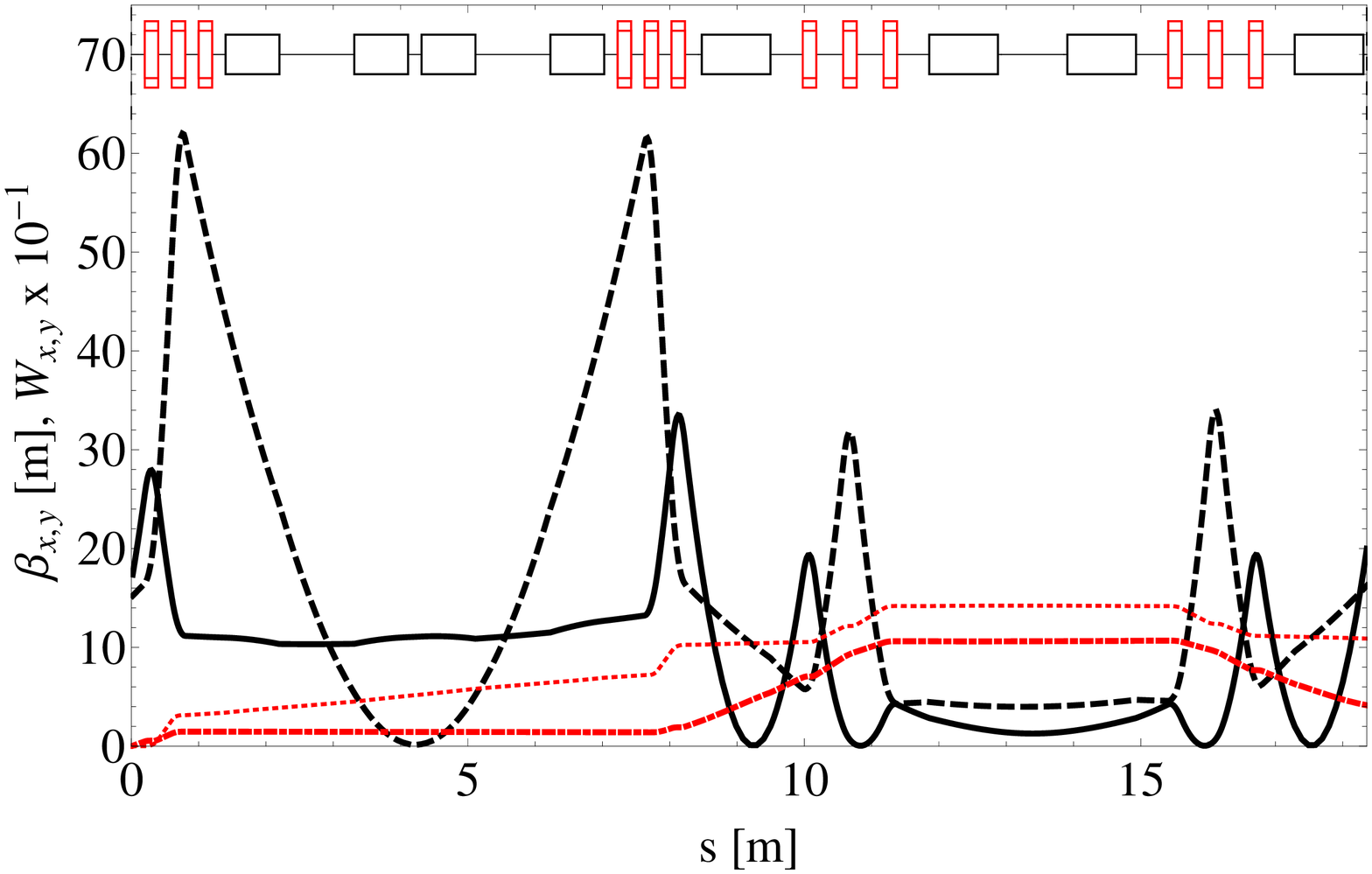}
  \caption{Matched $\beta$ functions in $x$ (solid, black) and $y$ (dashed, black) for the 1200~MeV path corrector in its maximum path length position, showing also the cumulative chromatic functions $W_x$ (heavy dots, red) and $W_y$ (light dots, red) from $s=0$ at the path corrector entrance.}
  \label{fig:betasmax1200}
\end{figure}

\begin{figure}[tbp]
  \centering
  \includegraphics[height=45mm]{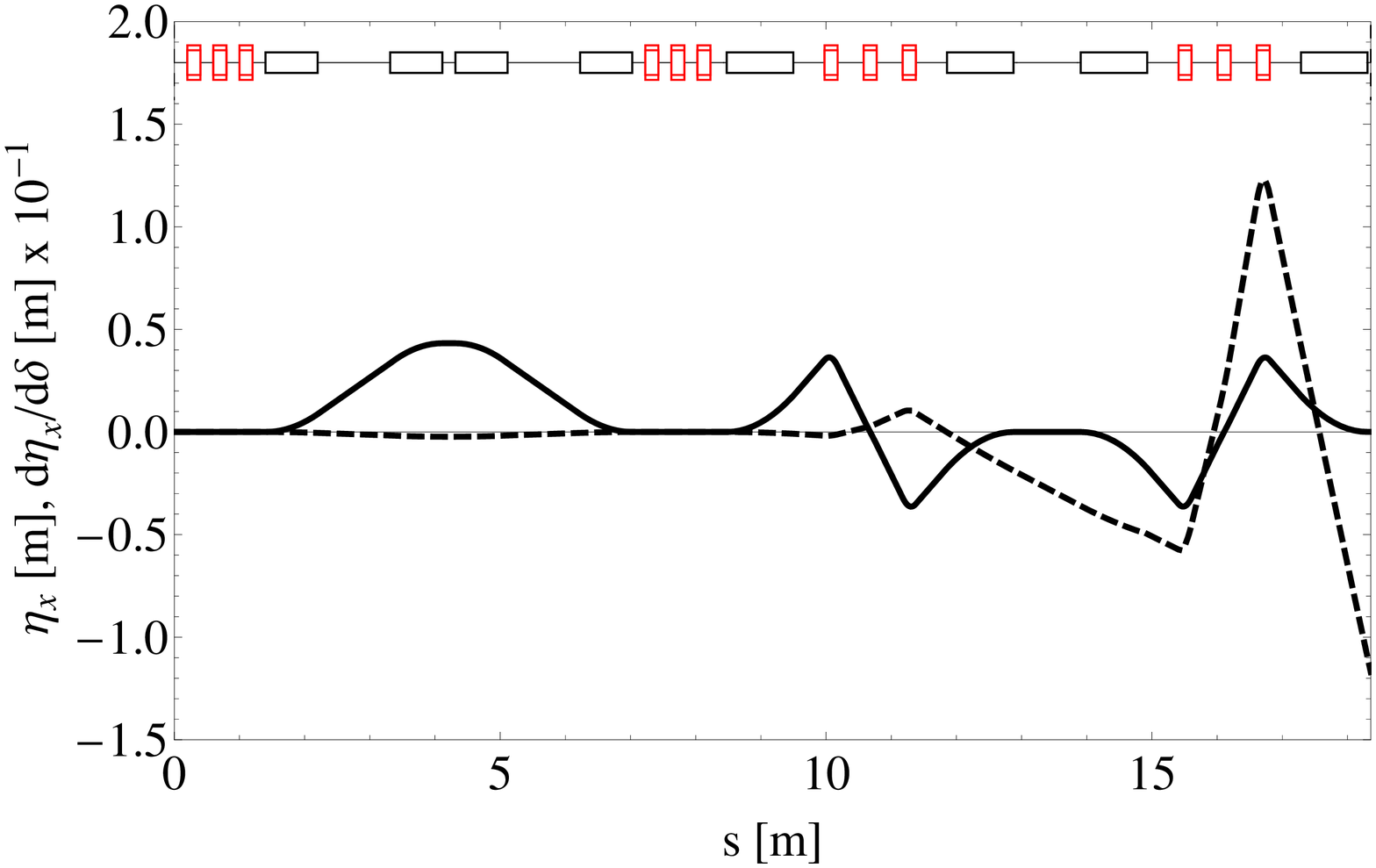}
  \caption{Dispersion $\eta_x$ and chromatic derivative of dispersion $d\eta_x /d\delta$ in the 1200~MeV path corrector in its maximum path length position. Similarly to the lower-energy case, sextupoles may be used to zero $d\eta_x /d\delta$ at the FC end.}
  \label{fig:disp1200max}
\end{figure}

Figure \ref{fig:r561200max} shows the development of $R_{56}$ and $T_{566}$ in the 1200~MeV path corrector in the maximum position, and Figures \ref{fig:betasmin1200} and \ref{fig:betasmax1200} show the change in Twiss functions as the FC is moved. Although the relative contribution to $R_{56}$ and $T_{566}$ from the FC is larger, its effect on the overall beam transport is modest. Similarly, the chromatic derivative of dispersion (shown in Figure \ref{fig:disp1200max}) is still manageable with local sextupole correction.

\section{\label{sec:conc}Discussion}
We have presented a dedicated beamline section that can perform path correction in a recirculating linac whilst leaving longitudinal dispersion unchanged, and that has little coupling to the downstream transverse optics. It may also be incorporated with an existing chicane-based bunch compression system to maintain any desired value of $R_{56}$. By decoupling the path length correction we propose that this scheme may be used more easily than the alternative method of perturbing arc optics, and we have shown that the focusing chicane required to do this can be kept to a length of $\sim$10~m at typical beam energies of c.~1~GeV. At this size the mechanical stability will be similar to that demonstrated on ALICE, giving confidence that a sub-1 degree accuracy of RF phase accuracy may be achieved. Changing nonlinearities are either small or readily correctable using standard techniques applied to linac beam transport, and may be applied during path correction changes using look-up tables if required. We propose that this technique is an attractive option for use in GeV-scale recirculating linacs.

\section{Acknowledgements}

We greatly appreciate the advice and discussions with David Douglas and Geoff Krafft, both of Thomas Jefferson National Laboratory.

\bibliographystyle{model1-num-names}







\end{document}